\documentclass[pre, superscriptaddress, secnumarabic, tightenlines,twocolumn,showpacs,showkeys,amsmath,amssymb,nofootinbib]{revtex4}

\usepackage[mathscr]{eucal} 
\usepackage{graphicx}

\allowdisplaybreaks[4]
\newcommand{\flow}[1]{\mathscr{#1}}
\newcommand{\vect}[1]{\pmb {#1}}
\newcommand{\tens}[1]{\pmb {\mathsf{#1}}}
\newcommand{\vi}{\pmb {\imath}}
\newcommand{\vecti}{\: {\bf{I}} \:}
\newcommand{\tscale}{\theta}
\newcommand{\Vscale}{\mathscr{V}}
\newcommand{\ximod}{{\xi}}
\newcommand{\sigmamod}{{\sigma}}

\newcommand{\ti}[1]{}  
\newcommand{\vol}[1]{\bf #1}  

\allowdisplaybreaks[4]
\begin{document}

\title{Jaynes' MaxEnt, Steady State Flow Systems and the \\ Maximum Entropy Production Principle}
\author{Robert K. Niven}
\email{r.niven@adfa.edu.au}
\affiliation{School of Engineering and Information Technology, The University of New South Wales at ADFA, Canberra, ACT, 2600, Australia.}
\date{6 August 2009}


\begin{abstract}
Jaynes' maximum entropy (MaxEnt) principle was recently used to give a conditional, local derivation of the ``maximum entropy production'' (MEP) principle, which states that a flow system with fixed flow(s) or gradient(s) will converge to a steady state of maximum production of thermodynamic entropy (R.K. Niven, Phys. Rev. E, in press). The analysis provides a steady state analog of the MaxEnt formulation of equilibrium thermodynamics, applicable to many complex flow systems at steady state. The present study examines the classification of physical systems, with emphasis on the choice of constraints in MaxEnt. The discussion clarifies the distinction between equilibrium, fluid flow, source/sink, flow/reactive and other systems, leading into an appraisal of the application of MaxEnt to steady state flow and reactive systems.

\end{abstract}


\maketitle

\section{\label{sect:intro}Introduction} 
%
Half a century ago, Jaynes established the maximum entropy (MaxEnt) principle as a method of inference for the solution of indeterminate problems of all kinds \cite{Jaynes_1957, Jaynes_1963, Jaynes_2003}, based on the relative entropy function (the negative Kullback-Leibler \cite{Kullback_L_1951} function):
\begin{equation}
\mathfrak{H}_{rel} = - \sum\limits_{i = 1}^s {p_i \ln \frac{{p_i }}{{q_i }}} 
\label{eq:relent}
\end{equation}
where $p_i$ is the probability of the $i$th distinguishable category or choice within a system, from $s$ such categories, and $q_i$ is the source or ``prior'' probability of category $i$. Maximization of \eqref{eq:relent}, subject to the constraints on a system, yields its ``least informative'' or ``maximally noncommittal'' probability distribution. Using the generic ``Jaynes relations'', this can then be used to infer the macroscopic properties of the system. 
Jaynes' method has been applied to the analysis of a vast number of phenomena throughout most fields of human study \citep[e.g.][]{Jaynes_2003, Kapur_K_1992}, and can be regarded as one of the most important discoveries of science.

Three decades ago, a new principle was proposed by Paltridge for the analysis of flow systems, the {\it maximum entropy production} (MEP) principle \cite{Paltridge_1975, Paltridge_1978}. This can be specified as ``{\it a flow system with many degrees of freedom, subject to fixed flows or gradients, will tend towards a steady state position of maximum production of thermodynamic entropy}''. Since its proposition, the MEP principle has been applied successfully to predict the steady states of a wide range of systems, including the Earth's climate system (oceans and atmosphere) \citep[e.g.][]{Paltridge_1975, Paltridge_1978, Kleidon_L_book_2005}; thermal (B\'enard) convection \citep{Ozawa_etal_2003}; material flow through ecological systems \citep{Meysman_B_2007} and biochemical processes \citep{Juretic_Z_2003, Dewar_etal_2006}. Lying outside present-day thermodynamics, the MEP principle appears to provide a unifying principle for the analysis of flow systems of all kinds.

Until recently, the theoretical basis of the MEP principle was unclear, limiting its acceptance. Philosophically, however, the MEP principle concerns reproducible behavior of a system, and therefore {\it must} be consistent with -- and hence derivable from -- Jaynes' MaxEnt method. The first approach in this direction was undertaken by Dewar \cite{Dewar_2003, Dewar_2005}, who examined a time-variant nonequilibrium system, using a MaxEnt analysis based on the probabilities of paths in phase space.  More recently, the author \cite{Niven_MEP} has given a local, conditional MaxEnt derivation of MEP, using the probabilities of instantaneous fluxes through each infinitesimal element of a flow system. 

The aim of this study is to explore the conceptual framework of the latter analysis in more detail.  Firstly, various types of physical systems are classified, to more fully understand {\it how} Jaynes' MaxEnt method can be applied, especially the handling of various types of constraints.  Several formulations of the MaxEnt method are then discussed, with application to equilibrium, flow and source/sink systems. The analysis leads into a discussion of Lagrangian multiplier constraints and the role of the potential function (generalized free energy concept) in MaxEnt analysis.

\section{\label{sect:sys}Types of Systems} 
%
Consider a physical system composed of discrete entities, each of which may adopt particular values of one or more physical quantities. 
In general, the system will be free to roam throughout its available ``category space''. In particular cases, however, the system will be constrained by mean values of particular physical quantities, whereupon the ``least informative'' or ``most probable'' state of the system can be inferred by Jaynes' MaxEnt method.

\begin{figure*}[t]
\setlength{\unitlength}{0.6pt}
  \begin{picture}(700,370)
   \put(20,0){\includegraphics[width=40mm]{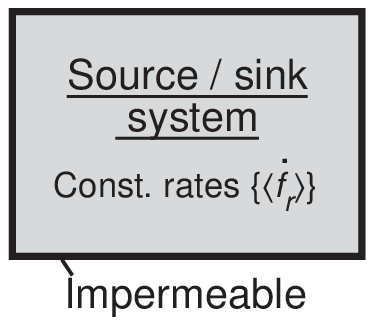} }
   \put(0,0){\small (c)}
   \put(320,0){\includegraphics[width=65mm]{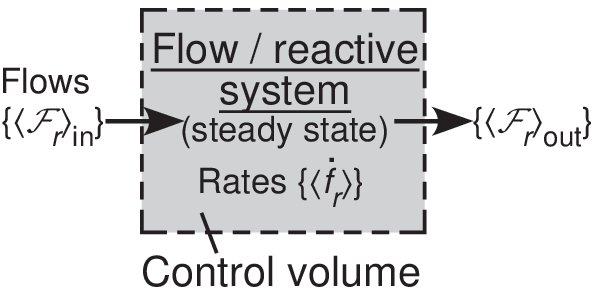} }
   \put(300,0){\small (d)}
   \put(20,190){\includegraphics[width=40mm]{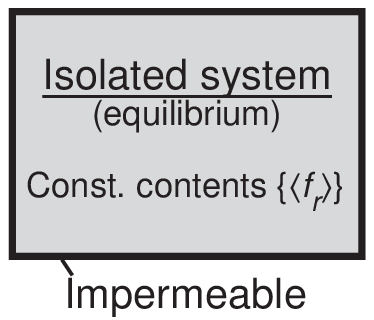}}
   \put(0,190){\small (a)}
   \put(320,190){\includegraphics[width=65mm]{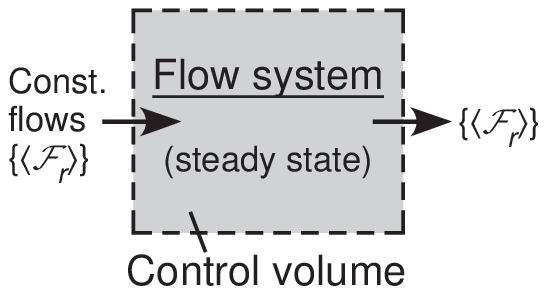}}
   \put(300,190){\small (b)}
  \end{picture}
\caption{Types of systems: (a) quantity-constrained (equilibrium) systems; (b) flow-constrained (steady state) systems; (c) rate-constrained (source-sink) systems; and (d) flow and reactive systems.}
\label{fig1}
\end{figure*}

With some thought, many physical systems can be classified into the following scheme, as illustrated in Figure 1:
\begin{list}{$\bullet$}{\topsep 1pt \itemsep 1pt \parsep 0pt \leftmargin 8pt \rightmargin 0pt \listparindent 0pt
\itemindent 0pt}
\item {\it Isolated systems} (Fig.\ 1a), consisting of a set of entities enclosed by some impermeable barrier, constrained by mean {\it contents} $\langle f_r \rangle$ of various physical parameters $f_r$, $r=1,...,R$.  If the content constraints are constant, such systems will converge towards an {\it equilibrium} position, which -- as is well known -- can be inferred by MaxEnt \cite{Jaynes_1957}.
\item {\it Flow systems} (Fig.\ 1b), defined by a {\it control volume}, a geometric region through which fluid(s) may flow (the Eulerian description), constrained by mean {\it flow rates} $\langle {\flow{F}_r} \rangle$ of various physical quantities $f_r$ through the system, $r=1,...,R$. If the flow rates are constant, such systems will converge towards a {\it steady state} position, which -- in principle -- can also be inferred by MaxEnt \cite{Niven_MEP}.
\item {\it Source/sink systems} (Fig.\ 1c), consisting of a set of entities enclosed by some impermeable barrier, constrained by mean {\it rates of change} $\langle \dot{f}_r \rangle$ of various physical parameters $f_r$ , $r=1,...,R$.  Such systems encompass chemical reaction kinetics, radioactive and biological growth/decay processes. If the rate constraints are constant, such systems will converge towards a {\it constant growth} position, which -- arguably -- can also be inferred by MaxEnt.   
\item {\it Flow and reactive systems} (Fig.\ 1d), again defined by a control volume, but in this case with different incoming flow rates $ \langle {\flow{F}_r} \rangle_{in}$, outgoing flow rates $\langle {\flow{F}_r} \rangle_{out}$ and rates of production $\langle {\dot{f}_r} \rangle$ of various physical quantities. If the flows and rate of production of each quantity are in balance, such that a system attains a steady state flow, it should be possible to infer this position by MaxEnt analysis.  
\item {\it Transient systems}, akin to those shown in Fig. 1d, but in general with time-varying incoming and/or outgoing flow rates and/or rates of production within the system. In general, such systems may not be amenable to analysis by MaxEnt, but in some special instances, they may be so amenable. Several such systems are examined by Dewar \cite{Dewar_2003, Dewar_2005} by a path MaxEnt analysis, and by Grandy \cite{Grandy_2008} using a quantum mechanics formulation.
\end{list}
Note that we here avoid the standard epithet ``{\it non-equilibrium system}'', which severely lacks precision; indeed, it is about as useful as the colour description ``non-blue''.

How can the MaxEnt method be applied to each above case? After a moment's reflection, it will be apparent that in isolated systems (Fig.\ 1a), the physical quantities $f_r$ within the mean constraints $\langle f_r \rangle$ are (usually) taken as {\it extensive} variables. By the zeroth law of thermodynamics, such properties should be uniformly distributed throughout the system at equilibrium. When applying MaxEnt to an isolated system, we are therefore justified -- in most circumstances -- to analyse the system on a whole-of-system basis.  This approach does not, however, apply in general, especially to the analysis of flow, source/sink and flow/reactive systems. In flow systems (Fig.\ 1b), for example, although the bulk flows into and out of the control volume may be specified around the control surface, many flow patterns could exist which are consistent with such boundary conditions. For maximum rigour, it is therefore necessary -- consistent with the formulation of other laws of fluid mechanics -- to apply MaxEnt to each infinitesimal element of the system. The analysis must therefore be conducted usings fluxes rather than flow rates\footnote{In engineering analysis, fluxes are vectors or tensors, expressed in SI units of quantities m$^{-2}$ s$^{-1}$, whereas flow rates are scalars, expressed in quantities s$^{-1}$.}. Similar considerations apply to source/sink, flow/reactive and transient systems, which also force the user to adopt a local rather than whole-of-system formulation of MaxEnt.

\section{\label{sect:MaxEnt}MaxEnt Analyses} 
We now turn to the mathematical treatment of each case.  In an {\it equilibrium system} (Fig. 1a), we consider the joint probability $p_{\vi}$ that an entity adopts particular values $f_{r{i_r}}$ of the physical parameters $f_r$, $r = \{1,...,R\}$. We therefore adopt the relative entropy:
\begin{equation}
\mathfrak{H}_{eq} = - \sum\limits_{\vi} {p_{\vi} \ln \frac{{p_{\vi} }}{{q_{\vi} }}} 
\label{eq:H_eq}
\end{equation}
where $i_r$ is the index of categories for the $r$th constraint, $\vi=\{i_1,...i_R\}$ and $q_{\vi}$ is the joint prior probability. In thermodynamics, the prior probabilities are typically handled in terms of the degeneracy $g_{{i_r}}=q_{{i_r}} G_r$ of each category ${i_r} \in \vi$, where $G_r=\sum\nolimits_{{i_r}} g_{{i_r}}$. Eq.\ \eqref{eq:H_eq} is subject to the natural and moment constraints:
\begin{gather}
\sum\limits_{\vi} {p_{\vi} }= 1, \qquad \sum\limits_{\vi} {p_{\vi} f_{r{i_r}} }= \langle {f_r } \rangle, \quad r = 1,...,R, 
\label{eq:C0Cr}
\end{gather}
where 
$\langle {f_r} \rangle$ is the mathematical expectation of $f_{r{i_r}}$.  Typical thermodynamic constraints include the mean internal energy $\langle U \rangle$, mean volume $\langle V \rangle$ and mean numbers of moles $\langle n_c \rangle$ of particles of each type $c$. Maximization of \eqref{eq:H_eq} subject to \eqref{eq:C0Cr} by Lagrange's method gives the inferred distribution, ``Jaynes' relations'', generalized Clausius equality and potential function of the system, as listed in Table 1.  We can further identify the dimensionless entropy and multipliers as functions of (historically) known variables; e.g.\ the thermodynamic entropy is $S=k \mathfrak{H}^*$, where $k$ is the Boltzmann constant, whilst the multipliers for the constraints $\langle U \rangle$, $\langle V \rangle$ and $\langle n_c \rangle$ are $\lambda_U = 1/kT$, $\lambda_V = P/kT$ and $\lambda_c = - \mu_c/kT$, where $T$ is the absolute temperature, $P$ is the absolute pressure and $\mu_c$ is the chemical potential of the $c$th constituent. In addition, the potential $\phi_{eq}$ equates to the Planck potential of the system, equivalent to the applicable free energy divided by $kT$ \cite{Niven_MEP}.  With these identifications, the relations in the left column of Table 1 provide the foundations of equilibrium thermodynamics \citep[e.g.][]{Callen_1960}.

\begin{table*}
\begin{tabular}{llll}
\hline
{\bf Property} &{\bf Equilibrium Systems} &{\bf Steady State Flow Systems} \\
\hline 
Categories & $f_{ri_r}$, for $i_r \in \vi$ & $\vect{j}_{ri_r}$, for $i_r \in \vecti$\\
\hline  
Probability  & $p_{\vi}=Prob(\{f_{r{i_r}} \}|B)  $ & $\pi_{\vecti}=Prob(\{\vect{j}_{r{i_r}} \}|B)$ \\
\hline  
Constraints  & $\langle 1 \rangle$ and $\{ \langle f_r \rangle \}$& $\langle 1 \rangle$ and $\{ \langle {\vect{j}}_r \rangle \}$\\
\hline 
Multipliers  & $\lambda_0$ and $\{ \lambda_r \}$ & $\zeta_0$ and $\{ \vect{\zeta}_r \}$\\
\hline 
Entropy  & $\mathfrak{H}_{eq} = - \sum\limits_{\vi} {p_{\vi} \ln \dfrac{{p_{\vi} }}{{q_{\vi} }}}$ 
 & $\mathfrak{H}_{st} = - \sum\limits_{\vecti} {\pi_{\vecti} \ln \dfrac{{\pi_{\vecti} }}{{\gamma_{\vecti} }}} $
 \\
\hline 
{Inferred distribution} & $p_{\vi}^* =  {Z}^{-1} q_{\vi} {\exp \bigl(- \sum\limits_{r = 1}^R  \lambda_r  f_{r{{i_r}}}} \bigr)$ 
&$\pi_{\vecti}^* =   {\flow{Z}}^{-1} { \gamma_{\vecti} \exp(  - \sum\limits_{r = 1}^R  \vect{\zeta}_r \cdot {\vect j}_{r{i_r}}   } )
$\\
& $Z =  e^{\lambda_0} = \sum\limits_{{\vi}} {q_{\vi} \exp \bigl(- \sum\limits_{r = 1}^R  \lambda_r  f_{r{{i_r}}}} \bigr)$ 
& $\flow{Z} =  e^{\zeta_0} = \sum\limits_{\vecti} \gamma_{\vecti} \exp(  - \sum\limits_{r = 1}^R  \vect{\zeta}_r  \cdot {\vect j}_{r{i_r}} )$ 
\\ 
\hline 
{Jaynes relations} & $\mathfrak{H}_{eq}^*  = \lambda_0  +  \sum\limits_{r = 1}^R \lambda_r \langle {f_r} \rangle $ 
& $\mathfrak{H}_{st}^*  = \zeta_0  +  \sum\limits_{r = 1}^R \vect{\zeta}_r \cdot \langle {\vect{j}_r} \rangle $
\\
& $\dfrac{{\partial \mathfrak{H}_{eq}^*}}{{\partial \langle {f_r } \rangle }} = \lambda_r $ 
& $\dfrac{{\partial \mathfrak{H}_{st}^*}}{{\partial \langle {\vect{j}_r} \rangle }} = \vect{\zeta}_r $
\vspace{4pt} 
\\
& $\dfrac{\partial ^2 \mathfrak{H}_{eq}^*}{\partial \langle {f_m} \rangle \partial \langle {f_r} \rangle} = \dfrac{\partial \lambda_r}{\partial \langle {f_m} \rangle}$ 
& $\dfrac{\partial ^2 \mathfrak{H}_{st}^*}{\partial \langle {\vect{j}_m} \rangle \partial \langle {\vect{j}_r} \rangle} = \dfrac{\partial \vect{\zeta}_r}{\partial \langle {\vect{j}_m} \rangle}$
\vspace{4pt} 
\\
& $\dfrac{\partial  \lambda_0 }{\partial \lambda_r}  = - \langle {f_r} \rangle $
 & $\dfrac{\partial  \zeta_0 }{\partial \vect{\zeta}_r}  = - \langle {\vect{j}_r} \rangle$
 \vspace{4pt} 
 \\
& $\dfrac{\partial^2 \lambda_0}{\partial \lambda_m \partial \lambda_r}   = - \dfrac {\partial \langle {f_r} \rangle}{\partial \lambda_m} $
 & $\dfrac{\partial^2 \zeta_0}{\partial \vect{\zeta}_m \partial \vect{\zeta}_r}  = - \dfrac {\partial \langle {\vect{j}_r} \rangle}{\partial \vect{\zeta}_m} $
 \vspace{2pt} 
 \\
\hline
Clausius &$d\mathfrak{H}_{eq}^* = \sum\limits_{r = 1}^R {\lambda _r \delta Q_r }$ 
& $d\mathfrak{H}_{st}^* = \sum\limits_{r = 1}^R {\vect{\zeta} _r \cdot \delta \vect{q}_r }$
\\
\hline
Potential function \hspace{8pt} &$d \phi_{eq} = - d \lambda_0 =  \sum\limits_{r = 1}^R {\lambda _r  \delta W_r }  +  \sum\limits_{r = 1}^R d \lambda_r \langle {f_r} \rangle$  \hspace{8pt}
& $d \phi_{st} = - d \zeta_0 =  \sum\limits_{r = 1}^R {\vect{\zeta} _r \cdot \delta \vect{w}_r }  +  \sum\limits_{r = 1}^R d \vect{\zeta}_r \cdot \langle {\vect{j}_r} \rangle$
\\
& $\hspace{15pt} =-d\mathfrak{H}_{eq}^*  + d \sum\limits_{r = 1}^R \lambda_r \langle {f_r} \rangle$ 
& $\hspace{15pt} =-d\mathfrak{H}_{st}^*  + d \sum\limits_{r = 1}^R \vect{\zeta}_r \cdot \langle {\vect{j}_r} \rangle$
\\
\hline
\end{tabular}
\caption{\label{table_eq}Assumptions, entropy function, inferred probability distribution and Jaynes relations, for (a) equilibrium and (b) steady state flow systems ($B$=background information, $\delta Q_r$ = $r$th generalized heat, $\delta W_r$ = $r$th generalized work, $\delta \vect{q}_r$ = $r$th flux of generalized heat, $\delta \vect{w}_r$ = $r$th flux of generalized work).} 
\end{table*}

We now consider {\it steady state systems}, as illustrated in Fig. 1b, for which -- as noted earlier -- it is necessary to apply MaxEnt to each infinitesimal element of the system. We thus consider the joint probability $\pi_{\vecti}$ that a volume element experiences instantaneous values ${\vect j}_{r{i_r}}$ of the fluxes of various quantities $f_r$ through the element. 
We therefore maximize the relative entropy:
\begin{equation}
\mathfrak{H}_{st} = - \sum\limits_{\vecti} {\pi_{\vecti} \ln \frac{{\pi_{\vecti} }}{{\gamma_{\vecti} }}} 
\label{eq:H_st}
\end{equation}
where $\vecti=\{i_1,...i_R\}$ and $\gamma_{\vecti}$ is the joint prior probability. $\mathfrak{H}_{st}^*$ can be termed the {\it flux entropy}, since it expresses the spread of the distribution of instantaneous local fluxes; it is fundamentally different to the thermodynamic entropy $S=k \mathfrak{H}_{eq}^*$. Eq.\ \eqref{eq:H_st} is subject to the natural and moment constraints:
\begin{gather}
\sum\limits_{\vecti} \pi_{\vecti}= 1, \qquad \sum\limits_{\vecti} \pi_{\vecti} {\vect j}_{r{i_r}} = \langle {\vect j}_r \rangle, \quad r = 1,...,R, 
\label{eq:C0Cr_st}
\end{gather}
Typical constraints of interest include the mean heat flux $\langle {\vect j}_{Q} \rangle$, fluid flux $\langle {\vect j}_{f} \rangle$, mass fluxes $\langle {\vect j}_{c} \rangle$ of each species $c$, momentum flux (stress tensor) $\langle {\tens{\tau}} \rangle$ and/or charge flux $\langle {\vect j}_{z} \rangle$. Application of MaxEnt then yields the inferred distribution and set of relations for a flow system at steady state, as listed in Table 1, where $\vect{\zeta}_r$ are the new (vector or tensor) Lagrangian multipliers, $\zeta_0$ is the Massieu function, $\flow{Z}$ is the partition function and ``$\cdot$'' represents the vector or tensor scalar product (as circumstances require).  

The flow system parameters listed in Table 1 can be further identified as functions of known parameters.  An important quantity of ``non-equilibrium thermodynamics'' is the entropy production, given for each infinitesimal element (per unit volume) by \cite{deGroot_M_1962, Prigogine_1967, 
Bird_etal_2006}:
\begin{gather}
\hat{\dot{\sigmamod}}= \sum\nolimits_{r=1}^R {\vect \nabla} Y_r \cdot \langle {\vect j}_r \rangle
\label{eq:sigma_dot_hat2}
\end{gather}
where $Y_r$ is the extensive variable associated with quantity $f_r$, whilst $\vect {\nabla}$ is the Cartesian gradient operator. 
Comparing \eqref{eq:sigma_dot_hat2} with the flux entropy at steady state, $\mathfrak{H}_{st}^*  = \zeta_0  +  \sum\nolimits_{r = 1}^R \vect{\zeta}_r \cdot \langle {\vect{j}_r} \rangle$, we see that both quantities contain a sum of scalar products of fluxes with other quantities. By monotonicity arguments for each $r$th pair of fluxes and gradients in \eqref{eq:sigma_dot_hat2}, making use of the fact that the fluxes are linearly independent, we can therefore identify the Lagrangian multipliers as:
\begin{equation}
{\vect{\zeta}}_r = - \frac {\tscale \Vscale}{k} {\vect \nabla} Y_r 
\label{eq:zetas}
\end{equation}
where $\tscale$ and $\Vscale$ are characteristic time and volume scales for the system. With these identifications, the relations on the right in Table 1 provide the foundation for the MaxEnt analysis of flow systems at steady state.

{\it Source-sink systems} can be analysed by MaxEnt in a similar manner.  Adopting the entropy \eqref{eq:H_st}, now based on the joint probability $\pi_{\vecti}$ of instantaneous local rates of production $\hat{\dot{\ximod}}_{\flow{L}_{d}}$ of quantities $d$, subject to constraints $\sum\nolimits_{\vecti} \pi_{\vecti}= 1$ and $\sum\nolimits_{\vecti} \pi_{\vecti} \hat{\dot{\ximod}}_{\flow{L}_{d}} = \langle \hat{\dot{\ximod}}_{d} \rangle$, one obtains relations similar to those listed in Table 1 for flow systems, but expressed in terms of $\langle \hat{\dot{\ximod}}_{d} \rangle$ rather than $\langle \vect{j}_r \rangle$. For chemical reaction systems, the Lagrangian multipliers conjugate to $\langle \hat{\dot{\ximod}}_{d} \rangle$ can be identified as $\zeta_d = (\tscale \Vscale/k) (A_d/T)$, where $A_d$ is the chemical affinity of reaction $d$. The local entropy production $\hat{\dot{\sigmamod}}=-\sum\nolimits_d \hat{\dot{\ximod}}_d A_d/T$ again emerges as an important quantity in this system.

For the MaxEnt analysis of simultaneously {\it flow and reactive systems} portrayed in Fig.\ 1d, at steady state, it is necessary to adopt composite constraints, given by one side of the balance equation for a conserved quantity:
\begin{equation}
\langle {\flow{F}_r} \rangle_{in} + \langle {\dot{f}_r} \rangle = \langle {\flow{F}_r} \rangle_{out}
\label{eq:cons}
\end{equation}
This new application of MaxEnt warrants further detailed examination.

\section{\label{sect:Mult}Multiplier Constraints and the MEP Principle} 
\begin{figure*}[t]
\setlength{\unitlength}{0.6pt}
  \begin{picture}(700,180)
   \put(40,10){\includegraphics[width=40mm]{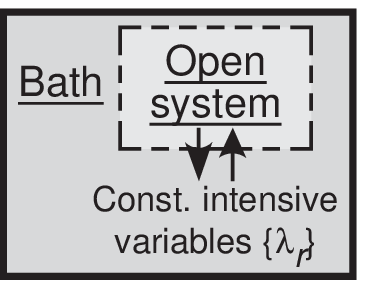}}
   \put(0,10){\small (a)}
   \put(320,0){\includegraphics[width=65mm]{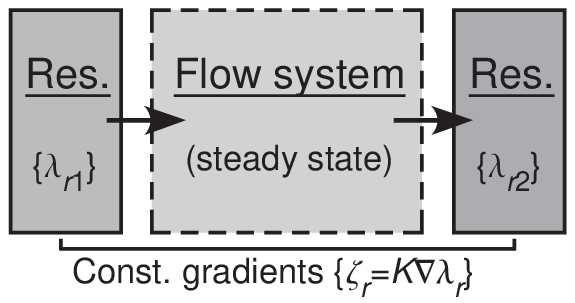}}
   \put(300,10){\small (b)}
  \end{picture}
\caption{Types of systems with multiplier ``constraints'': (a) intensive variable-constrained (open equilibrium) systems; and (b) gradient-constrained (steady state) systems.} 
\label{fig1}
\end{figure*}

A final point is to consider the alternative formulation of each of the systems discussed in \S\ref{sect:sys}-\ref{sect:MaxEnt}, in which the system is constrained -- not by mean values of physical variables -- but by their conjugate Lagrangian multipliers. Two examples are shown in Figure 2: 
\begin{list}{$\bullet$}{\topsep 1pt \itemsep 1pt \parsep 0pt \leftmargin 8pt \rightmargin 0pt \listparindent 0pt
\itemindent 0pt}
\item {\it Open equilibrium systems} (Fig. 2a), in which the equilibrium position is imposed by a surrounding bath of fixed intensive variables; and
\item {\it Gradient-constrained flow systems} (Fig. 2b), in which a steady state flow system is produced by the imposition of fixed gradients of intensive variables.
\end{list}
Irrespective of whether a set of physical quantities or their conjugate multipliers are adopted as constraints, the mathematical method for the analysis of each pair of systems is the same. This point was in fact appreciated by Jaynes \cite{Jaynes_1957}, but his discussion is rather oblique.  In addition to those shown in Figure 2, other multiplier-constrained systems are possible. {\it Hybrid equilibrium-steady state systems} must also be considered, by the imposition of fixed intensive variables of certain quantities, and gradients of other quantities; indeed, all combustion processes in the Earth's atmosphere may be conceptualized in this manner.

Since the work of Gibbs \cite{Gibbs_1875}, it has been standard practice in thermodynamics to use the free energy $F$ as the criterion for analysis of open equilibrium systems, wherein $F$ attains a minimum at equilibrium.  From a Jaynesian perspective, the free energy divided by $kT$ is equivalent to the negative Massieu function or {\it potential function} of a system: 
\begin{align}
\begin{split}
d \phi_{eq} &= - d \lambda_0 = d \Bigl( \frac{F}{kT} \Bigr) 
\\
&=-d \mathfrak{H}_{eq}^*  + d \sum\nolimits_{r = 1}^R \lambda_r \langle {f_r} \rangle 
\\
&=  - d\mathfrak{H}_{eq}^*  -d \mathfrak{H}_{ROU}
\label{eq:phi_eq}
\end{split}
\end{align}
where $\mathfrak{H}_{ROU}$ is the entropy of the rest of the universe. The potential $\phi_{eq}$ embodies the second law of thermodynamics, in the sense that spontaneous changes must take place by an interplay between changes in entropy within the system, $d \mathfrak{H}_{eq}^*$, and changes in entropy outside the system, $d \mathfrak{H}_{ROU}^*$, such that  $d \phi_{eq} \le 0$ \cite{Strong_H_1970, Craig_1988}.

In a similar vein, it is possible to adopt the potential function $\phi_{st}$ as the criterion for analysis of gradient-controlled flow systems. By a flow analog of the second law, $\phi_{st}$ will also attain a minimum at steady state. From its definition in Table 1, \eqref{eq:sigma_dot_hat2}, \eqref{eq:zetas} and the identification $\zeta_0 = - \phi_{st}$, the potential can be written:
\begin{equation}
d \phi_{st}  = - d \mathfrak{H}_{st}^* - \frac {\tscale \Vscale}{k} d \hat{\dot{\sigmamod}}
\label{eq:H_st2}
\end{equation}
Conditional on the assumption that $d \mathfrak{H}_{st}^* \ge 0$, we see that the minimum in $\phi_{st}$ will correspond to a state of maximum entropy production $\hat{\dot{\sigmamod}}$. The analysis therefore provides a local derivation of the MEP principle, conditional on a single assumption concerning the behavior of the flux entropy $\mathfrak{H}_{st}^*$. 

\section{\label{concl}Conclusions}
%
This study examines several kinds of physical system, including equilibrium, steady state flow, source/sink and flow/reactive systems. In these cases, it is shown that the ``stationary'' or ``constant'' position of the system can be inferred by Jaynes' MaxEnt, with the appropriate choice of relative entropy function and constraints. For equilibrium, flow and source/sink systems, constrained respectively by extensive variable contents, fluxes or rates of production, the Lagrangian multipliers can be identified respectively as functions of the intensive variables, gradients in the extensive variables or chemical affinity terms. In all cases, changes in the potential function (negative Massieu function) can be used as a criterion for stationarity, expressing the interplay between an increase in entropy (however defined) of a system, and an increase in entropy outside the system. For flow and source/sink systems, the minimum potential corresponds (conditionally) to a state of maximum entropy production. 

{\bf Acknowledgments}: The author thanks the European Commission for support as a Marie Curie Incoming International Fellow; UNSW and Technical University of Berlin for funding support; and the organizers and participants of MaxEnt09.

\end{document}